\documentclass[aps,prfluids,preprint,groupedaddress]{revtex4-2}

\usepackage{graphicx}
\usepackage{epstopdf, epsfig}
\usepackage{svg}
\usepackage{amsmath}
\usepackage{yhmath} 

\usepackage{amsmath,bm}
\usepackage{graphicx}
\usepackage[colorlinks=true, allcolors=blue]{hyperref}

\begin{document}
\title{Buckling and restabilization of a spinning elastic disk in a viscous fluid}
\author{Sifan Yin}
\affiliation{ Max Planck Institute for the Physics of Complex Systems, 01187 Dresden, Germany.}
\author{Paul R. Kaneelil}
\affiliation{ John A. Paulson School of Engineering and Applied Sciences, Harvard University, Cambridge, MA 02138, USA.}

\author{L.\ Mahadevan}
 \email{lmahadev@g.harvard.edu}
\affiliation{John A. Paulson School of Engineering and Applied Sciences, Harvard University, Cambridge, MA 02138, USA}
\affiliation{Department of Organismic and Evolutionary Biology, Harvard University, Cambridge, MA 02138}
\affiliation{Department of Physics, Harvard University, Cambridge, MA 02138}

\begin{abstract}
A thin elastic disk spinning at a constant speed in a viscous fluid is subject to two competing effects: rotation generates centrifugal tension that stabilizes the flat configuration, while the swirling flow entrained by the disk generates viscous shear that can destabilize it. Combining the Föppl–von Kármán equations for an elastic plate with the viscous tractions derived from the classical von Kármán swirling flow, we formulate an elastohydrodynamic stability problem. We show that this problem is mathematically equivalent to the shear-induced buckling of an accelerating disk, with the azimuthal viscous traction acting as an effective angular acceleration and the radial traction augmenting the centrifugal tension. Linear stability analysis characterizes the onset of buckling in terms of three dimensionless parameters: the ratio of the inner to outer radius, a solid number comparing centrifugal tension with bending stiffness, and a fluid number comparing the mass of fluid entrained in the boundary layer with that of the disk. For sufficiently compliant disks, increasing the angular velocity first destabilizes the flat configuration as viscous shear overcomes bending resistance, and subsequently restabilizes it as centrifugal tension grows faster than viscous shear. The resulting buckling modes exhibit a characteristic spiral morphology, with increasing azimuthal wavenumber and localization near the clamped inner edge as centrifugal tension increases. Our results delineate the conditions under which steady rotation can both induce and suppress an elastohydrodynamic instability.

\end{abstract}

\maketitle

\section{Introduction}
Rotation is a natural means of stiffening thin elastic structures. By generating tensile stresses, centrifugal forces suppress transverse deformations that would otherwise be resisted primarily by bending. This principle underlies the mechanics of rotating disks and membranes encountered in applications ranging from turbomachinery and data storage to deployable space structures\cite{hansen2001estimation,kim2000prediction}. However, when a compliant disk rotates in a viscous fluid, it also entrains the surrounding fluid, generating a swirling flow that exerts tangential tractions on its surfaces. While the radial component of these tractions augments the centrifugal tension, the azimuthal component generates in-plane shear that can destabilize the flat configuration. How does the competition between these effects determine the stability and shape of a spinning elastic disk?
 
The individual ingredients of this problem are classical. T. von K\'{a}rm\'{a}n separately studied the mechanics of thin elastic plates \citep{karman1907festigkeitsprobleme}, and the similarity solution associated with the swirling flow generated by an infinite rotating disk~\citep{karman1921laminare,cochran1934flow}. Rotating elastic disks have also been studied extensively in the context of centrifugal prestress \citep{stern1965rotationally,tang1970note,lamb1921vibrations}. More recently, studies of disks undergoing angular acceleration have shown that the associated in-plane shear stresses can give rise to buckling and wrinkling \citep{sader2019shear,coman2023shear}. In these systems, however, the destabilizing shear arises from changes in the rotation rate. In the presence of a viscous fluid, analogous shear stresses arise even when the disk rotates steadily.

Fluid–structure interactions in rotating elastic disks have also been studied in the context of membranes immersed in viscous fluids, where bending stiffness was neglected \citep{nowinski1984nonlinear}, and relatively stiff disks spinning in air, where coupling between transverse vibrations and aerodynamic pressure can lead to flutter
\cite{dAngelo1993natural,renshaw1994aerodynamically,kim2000prediction,hansen2001estimation,kang2004aeroelastic}. In the latter systems, the steady in-plane viscous tractions are generally too weak to cause appreciable deformation. Here, we consider a complementary regime: a thin, compliant disk rotating in a dense, viscous fluid, for which the steady tractions generated by the surrounding flow can themselves induce buckling. The instability of interest is thus a static, shear-induced instability in the co-rotating frame, rather than a dynamic flutter instability.  

\begin{figure}
\centering
\includegraphics[width=\textwidth]{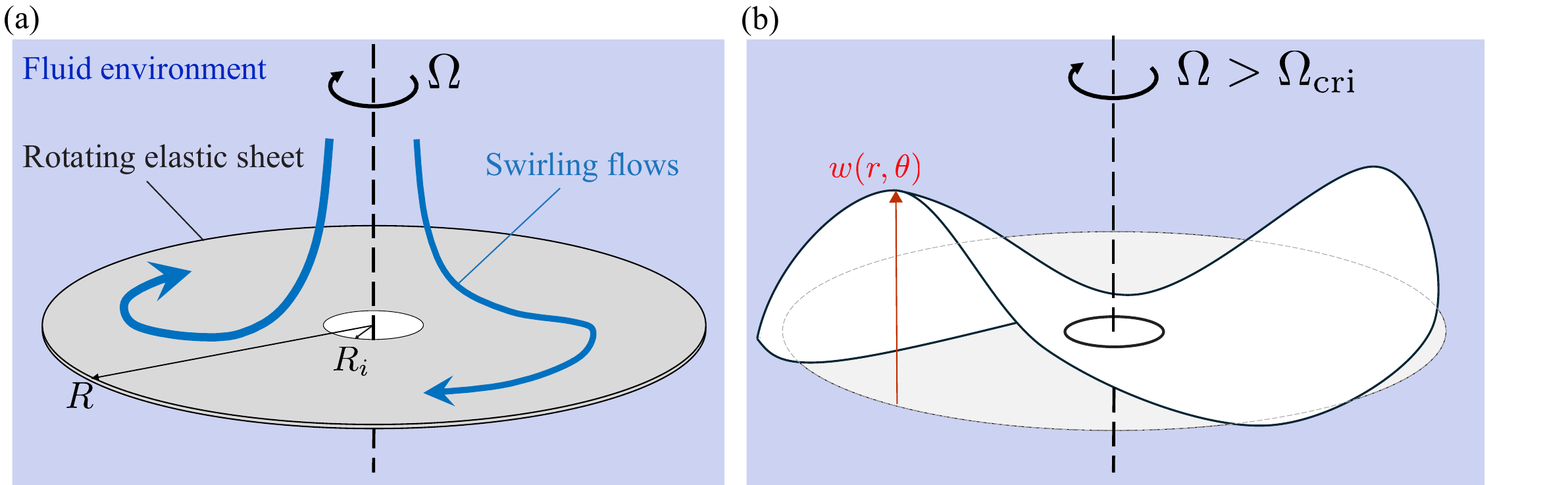}
\caption{ Schematic of a rotating elastic sheet submerged in a fluid environment with a constant angular velocity $\Omega$. This disk rotation can generate swirling flows on both faces, represented by blue lines. (a) An initially flat disk and (b) a buckled disk.} 
\label{fig1_schematic}
\end{figure}

A disk spinning at constant speed $\Omega$ entrains the surrounding fluid of viscosity $\nu$ into a von Kármán boundary layer of thickness $\delta=(\nu/\Omega)^{1/2}$ \cite{karman1921laminare,cochran1934flow}, which exerts tangential traction on both faces of the disk, as shown in Fig. ~\ref{fig1_schematic}. While the normal tractions on the two faces are balanced in the flat configuration, the radial and azimuthal tractions generate additional in-plane shear and tensile stresses that do not scale the same way with angular velocity. Thus, bending rigidity dominates at sufficiently low rotation rates, centrifugal tension dominates at sufficiently high rotation rates, and viscous shear can destabilize the disk at intermediate rotation rates. This difference in scaling suggests the possibility of reentrant stability, whereby a disk that buckles upon increasing its rotation rate recovers its flat configuration upon spinning faster.

To quantify this competition, we incorporate the stresses generated by the steady swirling flow into the linearized equations for a rotating elastic plate. The resulting formulation reveals a mathematical equivalence between the buckling of a steadily rotating disk in a viscous fluid and that of an accelerating disk in the absence of fluid tractions~\cite{sader2019shear}, with the important distinction that the destabilizing shear is now sustained rather than transient. This equivalence allows us to characterize the onset and morphology of the instability in terms of the disk geometry and the relative magnitudes of bending resistance, centrifugal tension and viscous shear.

We restrict our analysis to the onset of buckling under laminar base-flow conditions\cite{alfredsson2024flows} adopting a one-way coupling in which the flat rotating disk establishes the swirling flow, whose surface tractions determine the in-plane prestress of the disk. We neglect both the feedback of transverse deformations on the surrounding flow and the nonlinear post-buckling response, leaving these aspects for future study.

 In Section 2, we formulate the elastohydrodynamic equations in their dimensionless form. In Section 3, we determine the onset of buckling through linear stability analysis and compute the critical modes, and construct a stability phase diagram for disks of different geometries and material properties. We conclude in Section 4 with a discussion of the assumptions and   open questions for the future.

\section{Elastohydrodynamics of a spinning disk in viscous fluid}
\subsection{Geometry and governing equations}
Consider a thin elastic annular disk with outer radius $R$, inner radius $R_i$, thickness $h\ll R$, density $\rho_s$, Young's modulus $E$, and Poisson's ratio $\nu_p$. The disk is clamped to a spinning holder at its inner edge $r=R_i$ and free at the outer edge $r=R$, as shown in Fig.~\ref{fig1_schematic}(a). The initially flat disk is driven by the rotating holder at a constant angular velocity $\Omega$ about its axis and is surrounded by a viscous fluid of density $\rho_f$ and kinematic viscosity $\nu$. Above a critical angular velocity the flat disk buckles into a three-dimensional configuration, as shown in Fig.~\ref{fig1_schematic}(b). Let $w(r,\theta,t)$ denote the out-of-plane deflection of the disk mid-plane where $r$ and $\theta$ are the radial and azimuthal coordinates, and $t$ is the time. Then, the linearized {F\"{o}ppl--von~K\'{a}rm\'{a}n} dynamical equation for the transverse deflection of the plate reads as follows 
\begin{equation}\label{Eq_w_full}
  \rho_s h\frac{\partial^2 w}{\partial t^2}+ B \nabla^4 w - h \pmb{\nabla} \cdot (\pmb{\sigma} \cdot \pmb{\nabla} w)=0,
\end{equation}
where $B=Eh^3/[12(1-\nu_p^2)]$ is the flexural rigidity, $\pmb \sigma$ is the in-plane stress tensor. {Because the swirling flows are symmetric for the upper and lower surfaces, the flow-generated normal forces are balanced.}\par

The in-plane stress tensor $\pmb{\sigma}$ is balanced by the centrifugal forces ${\bm T}^{\rm c}=h\rho_s\Omega^2r{\bm e}_r$ arising from steady rotation, and the viscous traction forces per unit area ${\bm T}^{\rm fl}=T_r{\bm e}_r+T_\theta{\bm e}_\theta$ exerted by the surrounding swirling flow on the top and bottom surfaces, which are approximated as the von K\'{a}rm\'{a}n similarity solution \citep{karman1921laminare,cochran1934flow}
\begin{equation}
     T_r = F'\left(0\right)\rho_f(\Omega^3\nu)^{1/2}r,
     ~{\rm and}~
     T_{\theta} = -G'\left(0\right) \rho_f(\Omega^3\nu)^{1/2}r,
\end{equation}
with $G'\left(0\right) \approx 0.616$ and $F'\left(0\right)\approx 0.510$.

The in-plane force balance equation of the mid-plane is
\begin{equation}\label{Eq_in_plane_force_balance}
h\nabla\cdot{\bm \sigma}+2{\bm T}^{\rm fl}+{\bm T}^{\rm c}=0.
\end{equation}

Inserting $T_\theta$ to Eq.~(\ref{Eq_in_plane_force_balance}) and combining the free traction boundary condition $\sigma_{r\theta}|_{r=R}=0$, we solve the shear stress of the disk
\begin{equation}
    \sigma_{r\theta}=-\frac{G'\left(0\right) \rho_f(\Omega^3\nu)^{1/2}}{2h}\frac{R^4-r^4}{r^2}.
\end{equation}
When neglecting the swirling flow-generated surface traction forces, i.e., $T_r=T_\theta=0$, the centrifugal forces-induced stresses $\sigma_{rr}^{\rm c}$ and $\sigma_{\theta\theta}^{\rm c}$ are both tensile, thus stabilizing the flat configuration against the out-of-plane buckling, and are given as \citep{stern1965rotationally,love2013treatise,sader2019shear} 
\begin{subequations}
    \label{eq:stress}
    \begin{align}
       &\sigma_{rr}^{\rm c} = \frac{\rho_s\Omega^2 \left(R^2-r^2\right) \left\{ \left(3+\nu_{\rm p}\right)\left[\left(1+\nu_{\rm p}\right)r^2+\alpha^2\left(1-\nu_{\rm p}\right)\left(R^2+r^2\right)\right]-\alpha^4\left(1-\nu_{\rm p}^2\right)R^2     \right\}}{8\left[1+\nu_{\rm p}+\alpha^2\left(1-\nu_{\rm p}\right)\right]r^2},\\
       &\sigma_{\theta\theta}^{\rm c}=\frac{\rho_s\Omega^2}{8\left[1+\nu_{\rm p}+\alpha^2\left(1-\nu_{\rm p}\right)\right]r^2}\left\{\alpha^4 R^2\left(1-\nu_{\rm p}^2\right)\left(R^2+r^2\right)\right.
        \notag\\
        &~~~\left. -\alpha^2\left(1-\nu_{\rm p}\right)\left[(3+\nu_{\rm p})R^4+\left(1+3\nu_{\rm p}\right)r^4\right]+\left(1+\nu_{\rm p}\right)\left[(3+\nu_{\rm p})R^2-\left(1+3\nu_{\rm p}\right)r^2\right]r^2\right\},
    \end{align}
\end{subequations}
where $\alpha=R_i/R$ is the ratio of inner to outer radius. \par
With nonzero $T_r$, the normal stresses are modified to 
\begin{equation}\label{Eq_Coef_Lambda}
    \sigma_{rr} = \left(1+\Lambda\right)\sigma_{rr}^{\rm c}, ~{\rm and}~
    \sigma_{\theta\theta} = \left(1+\Lambda\right)\sigma_{\theta\theta}^{\rm c},
\end{equation}
where the coefficient $\Lambda = 2F'\left(0\right)\rho_f \nu^{1/2}/\left(h\rho_s\Omega^{1/2}\right)$. 
This means the radial surface traction amplifies the tensile stresses generated by centrifugal forces.
The instability therefore arises from the competition between two prestress fields: a stabilizing tension and a destabilizing viscous shear.

\subsection{Boundary conditions and  non-dimensionalization}
The boundary conditions for $w(r,\theta)$ prescribe clamped conditions at the inner edge $r=R_i$ and a stress free condition at the outer edge $r=R$
\begin{subequations}\label{Eq_BC_full}
\begin{gather}
w|_{r=R_i} = 0,~
\frac{\partial w}{\partial r} \bigg|_{r=R_i}=0,
\\
\left[\frac{\partial^2 w}{\partial r^2}+ \frac{\nu_p}{r}\frac{\partial w}{\partial r}+\frac{\nu_p}{r^2}\frac{\partial^2 w}{\partial \theta^2} \right]\bigg|_{r=R} = 0,
\\
\left[\frac{\partial^3 w}{\partial r^3} + \frac{1}{r} \frac{\partial^2 w}{\partial r^2} - \frac{1}{r^2} \frac{\partial w}{\partial r} - \frac{\left(3-\nu_p\right)}{r^3} \frac{\partial^2 w}{\partial \theta^2} + \frac{\left(2-\nu_p\right)}{r^2} \frac{\partial^3 w}{\partial r \partial \theta^2} \right]\bigg|_{r=R} = 0.
\end{gather}
\end{subequations}
Periodicity is imposed in the azimuthal direction $\theta$. \par

The governing equations are non-dimensionalized by defining the dimensionless variables $\tilde{w} = w/h$, $\tilde{r} = r/R$, $\varepsilon=h/R$, $\tilde t=t\Omega$, and $\pmb{\tilde {\sigma}}=\pmb{\sigma}/[\rho_s\Omega^2 R^2]$. Substituting Eq.~(\ref{eq:stress}) and the dimensionless variables into Eq.(\ref{Eq_w_full}) gives the dimensionless partial differential equation (dropping tildes),

\begin{equation}\label{Eq_w_nondimen_equation}
\begin{aligned}
   &\frac{\partial^2 w}{\partial  t^2}=\\
   &-\frac{1}{So}\bigg[ 
    \frac{\partial^4 w}{\partial r^4}+ \frac{2}{r}\frac{\partial^3 w}{\partial r^3} +\frac{1}{r^2}\bigg(2\frac{\partial^4 w}{\partial r^2 \partial \theta^2} -\frac{\partial^2w}{\partial r^2}\bigg) +
    \frac{1}{r^3}\bigg( \frac{\partial w}{\partial r} -2\frac{\partial^3 w}{\partial r \partial \theta^2}  \bigg)  +\frac{1}{r^4}\bigg(4\frac{\partial^2w}{\partial \theta^2}+\frac{\partial^4 w}{\partial \theta^4}\bigg) \bigg] \\ 
&+\left(1+2F'(0)Fl\right) \left[ \left(-A_1r^2+\frac{B_1}{r^{2}}+C_1\right)\frac{\partial^2w}{\partial r^2} + \left(-3A_1r-\frac{B_1}{r^{3}}+\frac{C_1}{r}\right) \frac{\partial w}{\partial r} + \left(-A_2-\frac{B_1}{r^{4}}+\frac{C_1}{r^{2}}\right)\frac{\partial^2w}{\partial\theta^2}\right] \\ 
    &+G'(0)Fl\left(\frac{r^4-1}{r^3}\frac{\partial^2 w}{\partial r\partial \theta}+\frac{r^4+1}{r^4}\frac{\partial w}{\partial \theta}\right),
\end{aligned}
\end{equation}
where the expressions for the algebraic coefficients $A_{1,2}$, $B_1$ and $C_1$ are given in the Supplementary Material [SM Eq.~(S7)]. The dimensionless boundary conditions at $r=\alpha$ and $r=1$ are of the same form as shown in Eq.~(\ref{Eq_BC_full}).

\par
Our model is mathematically equivalent to the spin-up disk problem \cite{sader2019shear,coman2023shear}. We have the effective angular velocity and angular acceleration
\begin{equation}\label{Eq_Comp_Sader}
        \bar\omega^2 = So\left[1+2F'(0)Fl\right],\quad {\rm and}\quad
        \bar\Omega=2SoG'(0)Fl,
\end{equation} 
defined in Eq.(11) in \cite{sader2019shear}.
When there is no fluid environment, i.e., $Fl=0$, the soft disk rotates at a constant angular velocity, $\bar\omega^2=So$, and $\bar\Omega=0$. Three dimensionless parameters govern the problem: the radius ratio, $\alpha$, the solid number, $So$, defined as the ratio of centrifugal tension to bending stiffness, and the fluid number, $Fl$, which compares the mass of fluid entrained in the boundary layer with that of the disk, or equivalently the viscous traction with the centrifugal body force.
\begin{equation}
   So = \frac{12(1-\nu_p^2)\rho_s \Omega^2 R^4}{Eh^2} ~~\textrm{and}~~Fl =  \frac{\rho_f}{h\rho_s}\left(\frac{\nu}{\Omega}\right)^{1/2}.
  \label{eq:dimlessparams}
\end{equation}
With these definitions, the coefficient $\Lambda$ in Eq.~(\ref{Eq_Coef_Lambda}) recasts
$\Lambda = 2F'\left(0\right)Fl$.


\section{Onset of buckling instability}
\subsection{Linear stability analysis}
We consider the onset of disk buckling in a steady state relative to the co-rotational frame, so that the normal displacement of the buckled disk is time-independent, $w=w(r,\theta)$, and the inertial term $\partial^2 w /\partial t^2$ in Eq.~(\ref{Eq_w_nondimen_equation}) vanishes. We seek normal modes of the form $w(r,\theta)={\rm Real}\left[\phi(r)e^{{\rm i}m\theta}\right]$ where $m=0,1,2,...$ and $\phi(r)$ is a complex function of $r$. Substituting this ansatz into the governing equation Eq.~(\ref{Eq_w_nondimen_equation}) and boundary conditions Eq.~(\ref{Eq_BC_full}) yields a fourth order eigenvalue problem for $\phi(r)$ of the form
\begin{equation}\label{Eq_phi_ODE}
    {\mathcal A}(m,{So})\phi(r) = Fl{\mathcal B}(m)\phi(r),
\end{equation}
where we define the operators  
\begin{equation}
\begin{aligned}
    &{\mathcal D}_m=\left[\frac{{\rm d}^4}{{\rm d} r^4}+\frac{2}{r}\frac{{\rm d}^3}{{\rm d}r^3}- \frac{2m^2+1}{r^2}\frac{{\rm d}^2}{{\rm d}r^2}
    +\frac{2m^2+1}{r^3}\frac{{\rm d}}{{\rm d}r}+\frac{m^2(m^2-4)}{r^4}\right]\\
    &{\mathcal N}_m=\left[\left(-A_1r^2+\frac{B_1}{r^2}+C_1\right)\frac{{\rm d}^2}{{\rm d}r^2} +\left(-3A_1r-\frac{B_1}{r^3}+\frac{C_1}{r}\right)\frac{{\rm d}}{{\rm d}r}-m^2\left(-A_2-\frac{B_1}{r^4}+\frac{C_1}{r^2}\right)\right].
    \end{aligned}
\end{equation}
\begin{equation}
    {\mathcal A}(m,{So}) = \frac{1}{So}{\mathcal D}_m-{\mathcal N}_m
\end{equation}
\begin{equation}
    {\mathcal B}(m)=2F'(0){\mathcal N}_m+{i}{m}G'(0)\left(\frac{r^4-1}{r^3}\frac{\rm d}{{\rm d} r}+\frac{r^4+1}{r^4}\right),
\end{equation}
and the eigenvalue $Fl$ is to be determined.
The associated boundary conditions are
\begin{subequations}\label{Eq_phiBCs}
\begin{gather}
\phi|_{r=\alpha} = 0, 
\phi^{(1)} |_{r=\alpha} =0,
\\
\bigg[\phi^{(2)}+ b_1 \phi^{(1)} + b_0 \phi \bigg]\bigg|_{r=1} = 0,
\bigg[\phi^{(3)} + a_2 \phi^{(2)} + a_1 \phi^{(1)} + a_0 \phi \bigg]\bigg|_{r=1} = 0,
\end{gather}
\end{subequations}
and the coefficients $a_i$ and $b_i$ are given in Eq. (S13) in SM. 
\par
\begin{figure}
    \centering\includegraphics[width=\linewidth]{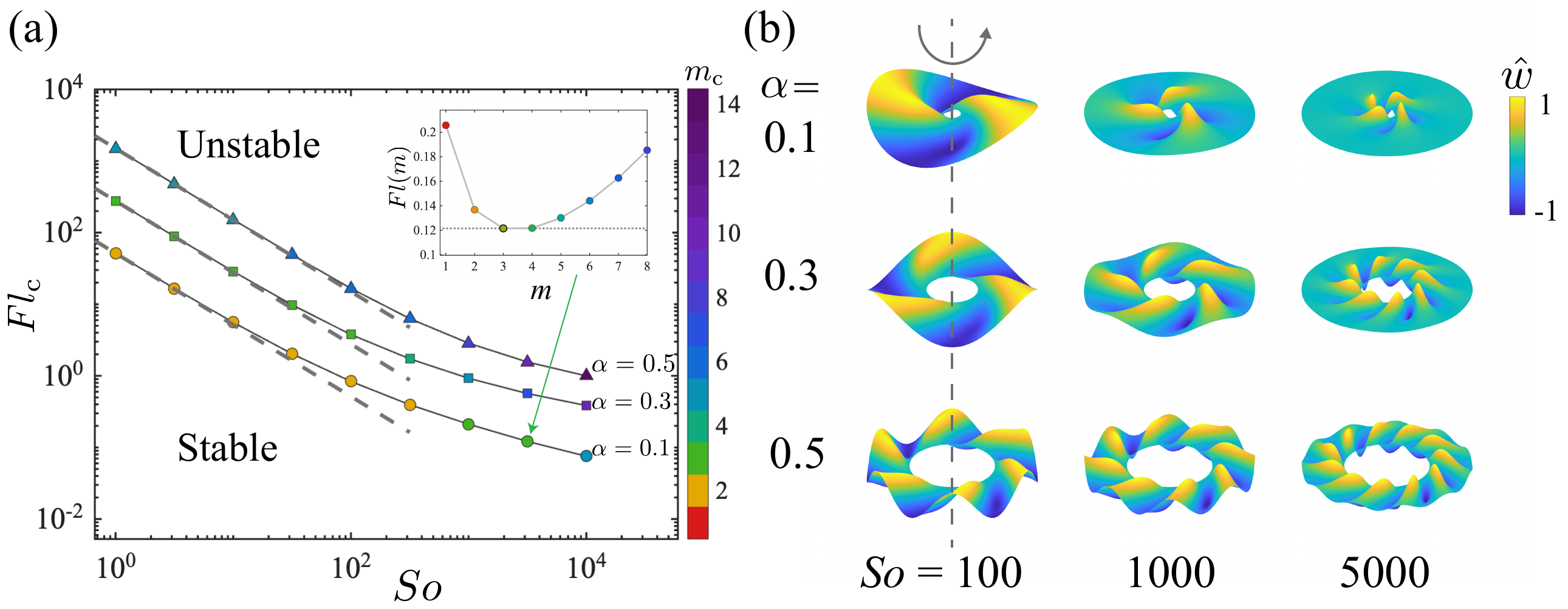}
    \caption{ (a) Critical fluid number $Fl_{\rm c}$ solved by Eq.~(\ref{Eq_phi_ODE}) as a function of the solid number $So$ for $\alpha=0.1$ (circles), $0.3$ (squares) and $0.5$ (triangles); the marker color gives the critical mode number $m_{\rm c}$. The disk is unstable for $Fl>Fl_{\rm c}$. Inset: $Fl(m)$ for $\alpha=0.1$ and $So\approx3.2\times10^3$; the minimum gives critical buckling mode $m_{\rm c}=3$. Dashed lines represent the bending-dominated limit $Fl_{\rm c}\propto 1/So$ when $So$ is small. (b) Critical buckling modes for $\alpha=0.1$, $0.3$, $0.5$ and $So=100$, $1000$, $5000$; color represents the normalized normal displacement $\hat w=w/\max(|w|)$. All results for $\nu_p=1/4$.}\label{fig2_phase_diagram_simulation}
\end{figure}

 The eigenvalue problem Eq.~(\ref{Eq_phi_ODE}--\ref{Eq_phiBCs}) is solved 
using a compound matrix method \citep{fu2011compound}. Fixing $So$, we search the smallest real $Fl$ with varying mode number $m$, leading to the results shown in Fig.~\ref{fig2_phase_diagram_simulation}(a) inset. The smallest $Fl_{\rm c}$ and its corresponding $m_{\rm c}$ are the critical values denoting the onset of buckling. The corresponding  eigenfunctions $\phi_{\rm c}(r)=f_{\rm c}(r)+ i g_{\rm c}(r) $ define the critical buckling mode. Then, on expanding the real and imaginary parts of the critical functions, $f_{\rm c}(r)$ and $g_{\rm c}(r)$ as a superposition of Chebyshev polynomials, allows us to write the buckling onset deflection as $w_{\rm c}(r,\theta)=f_{\rm c}(r){\rm cos}(m_{\rm c}\theta)+g_{\rm c}(r){\rm sin}(m_{\rm c}\theta)$.\par 


\subsection{Results}
Fig.~\ref{fig2_phase_diagram_simulation}(a) shows the critical fluid number $Fl_{\rm c}$ as a function of the solid number $So$ for inner radii $\alpha=0.1$, $0.3$ and $0.5$, with the color of each marker giving the critical mode number $m_{\rm c}$. For a fixed $So$, the disk is unstable when $Fl>Fl_{\rm c}$ reflecting the fact that the swirling-flow-induced shear stress destabilizes the disk, whereas the centrifugal tension, together with the radial tension due to the outward viscous drag, stabilizes it. 

Over four decades of $So$, $Fl_{\rm c}$ decreases by nearly three orders of magnitude, and two regimes can be distinguished. At small $So$, centrifugal tension is negligible and the fluid-induced stresses are resisted by bending alone. The eigenvalue problem Eq. ~(\ref{Eq_phi_ODE}) recasts $\mathcal D_m \phi(r) = So Fl \mathcal B\left(m\right)\phi(r)$, so that $So$ and $Fl$ enter only through their product $So\,Fl$, which measures the fluid shear relative to the bending stiffness. The threshold therefore reduces to $Fl_{\rm c}\simeq s^*(\alpha)/So$, with $s^*\approx 51$, $277$ and $1.5\times10^3$ for $\alpha=0.1$, $0.3$ and $0.5$, respectively, as shown in Fig.~\ref{fig1_schematic}(a) dashed lines.\par
At large $So$, bending becomes negligible and stability is controlled by the ratio of shear to tension, $\lambda=G'(0)Fl/[1+2F'(0)Fl]$, which does not depend on $So$. In this membrane limit, buckling requires a compressive principal stress, which first appears at the clamped edge when $\lambda$ exceeds $\lambda_{\rm b}(\alpha)$ [see Eq.~(S19) in SM]. Therefore $Fl_{\rm c}\to Fl_{\rm c}^{\rm lim}=\lambda_{\rm b}/[G'(0)-2F'(0)\lambda_{\rm b}]$, with $Fl_{\rm c}^{\rm lim}\approx0.011$, $0.092$ and $0.24$ for $\alpha=0.1$, $0.3$ and $0.5$. Because the drag that generates the shear also adds radial tension, $\lambda$ is bounded by $G'(0)/\left[2F'(0)\right]$; a finite $Fl_{\rm c}^{\rm lim}$ thus exists only when $\lambda_{\rm low}<G'(0)/[2F'(0)]\approx0.60$, which holds for all radius ratios considered here. The limit is approached slowly, so that $Fl_{\rm c}$ is still well above $Fl_{\rm c}^{\rm lim}$ at $So=10^4$, and the crossover between the two regimes appears as the bend of the curves around $So\sim10^2$--$10^3$. $Fl_{\rm c}$ increases with $\alpha$, because a narrower annulus is stiffer in bending. The critical mode number increases with both $So$ and $\alpha$, from $m_{\rm c}=2$ to $5$ for $\alpha=0.1$; from $3$ to $10$ for $\alpha=0.3$; and from $5$ to $14$ for $\alpha=0.5$ over the range shown. The inset shows $Fl(m)$, the smallest critical fluid number of each azimuthal mode, for $\alpha=0.1$ and $So\approx3.2\times10^3$. The minimum, $Fl_{\rm c}=0.121$, is attained at $m_{\rm c}=3$, while the value of $Fl$ at mode $m=4$ is only slightly larger than $Fl_{\rm c}$. This flat minimum is characteristic of large $So$, where neighboring modes become nearly degenerate, so that the mode selected in practice may be sensitive to imperfections, as also noted for the acceleration-induced problem \citep{sader2019shear}.

Typical critical buckling modes for different $\alpha$ and $So$ are shown in Fig.~\ref{fig2_phase_diagram_simulation}(b), where color represents the normalized normal displacement $\hat w=w/{\rm max}(|w|)$. At $So=100$ the deflection spans the whole annulus, whereas with increasing $So$ it becomes confined to the vicinity of the clamped inner edge, leaving the outer part of the disk nearly flat, and $m_{\rm c}$ increases accordingly. This confinement reflects the distribution of the prestress, indicating that the normalized shear stress $\sigma_{r\theta}\propto(1-r^4)/r^2$ is largest at the inner edge and vanishes at the free edge, while the centrifugal tension, which grows faster with rotation rate than the viscous shear, stiffens the outer region. The same mechanism confines the buckled modes in the acceleration-induced problem \citep{sader2019shear,coman2023shear}. For a narrow annulus ($\alpha=0.5$) the deformation fills the whole radial width $1-\alpha$, so the azimuthal wavelength is set by the width and the mode number is high even at small $So$. All modes display a characteristic spiral twist, the hallmark of shear-induced buckling.\par

\begin{figure}
    \centering
    \includegraphics[width=0.8\linewidth]{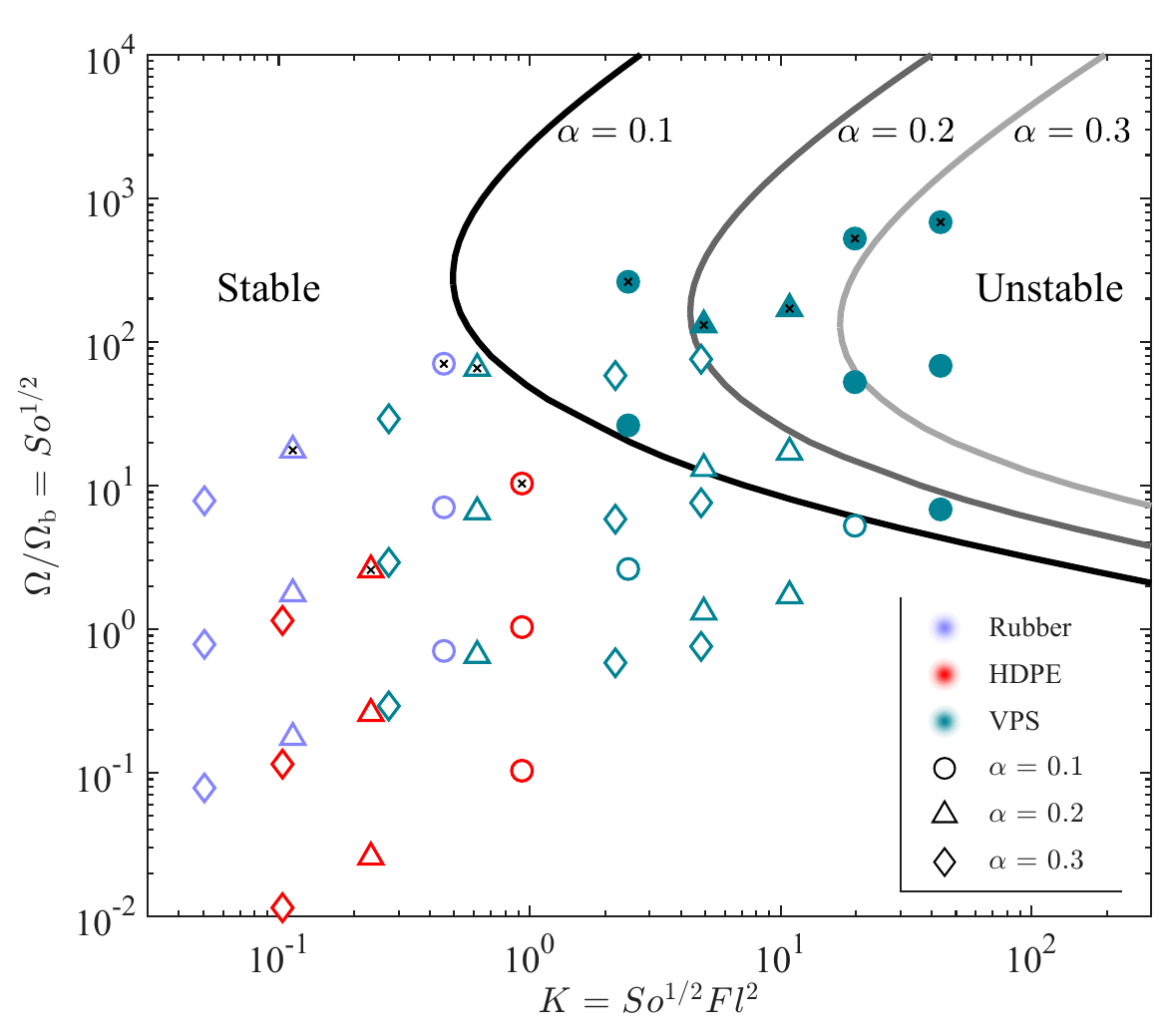}
    \caption{Stability phase diagram spanned by $K=So^{1/2}Fl^2$ and $\Omega/\Omega_{\rm b}=So^{1/2}$. Each C-shaped curve is mapped from $Fl_{\rm c}(So)$ with $\nu=0.5$ and $\alpha=0,1$, $0.2$, ~$0.3$. The right regions enclosed by the curves are unstable. The lower branch represents the onset of buckling and the upper branch represents a re-stabilization with increasing angular velocity. Colors denote three materials from ref.~\cite{kaneelil2026}: the natural rubber (blue), the high-density polyethylene (HDPE, red), and the vinyl polysiloxane (VPS, green). Circles, triangles and diamonds denote different radius ratios. Filled (open) markers correspond to unstable (stable) states. Small crosses mark $Re=\Omega R^2/\nu>2.6\times10^5$, beyond the laminar regime of the von K\'arm\'an flow \cite{lingwood1995absolute}. The physical parameters of different soft disks are given in Table.~S1.}
    \label{fig3_Omega_cr}
\end{figure}

\subsection{Estimates of parameters for representative geometries and materials}
Fig.~\ref{fig2_phase_diagram_simulation}(a) does not directly predict when a given disk buckles, because both control parameters vary with the rotation rate, $So\propto\Omega^2$ and $Fl\propto\Omega^{-1/2}$. Spinning up a disk of fixed material and geometry in a given fluid follows a path of constant 
\begin{equation}
K \equiv So^{1/2}Fl^2=\sqrt{\frac{12\left(1-{\nu_p}^2\right)}{E}} \frac{\rho_f^2 \nu R^2}{\rho_s^{3/2}h^3}.
\end{equation}
We therefore represent the stability boundary in the $( K,\Omega/\Omega_b)$ plane (Fig.~\ref{fig3_Omega_cr}), where $ \Omega_{\rm b} = \frac{h}{R^2}\sqrt{\frac{E}{12\left(1-\nu_p^2\right)\rho_s}}$ and $\Omega/\Omega_{\rm b}=So^{1/2}$. \par
The stability boundary is C-shaped for every $\alpha$, which is obtained by mapping the $Fl_{\rm c}(So)$ curve from Fig.~\ref{fig2_phase_diagram_simulation}(a) with $\nu=0.5$. The lowest value $K=K_{\min}$ sets the stable and unstable regions. When $K<K_{\rm min}$, the disk remains flat at any rotation rate; whereas a disk with $K>K_{\min}$ buckles and recovers to a stable state with increasing $\Omega$. This re-stabilization reflects the faster growth of centrifugal tension $\propto\Omega^2$, than of viscous shear $\propto\Omega^{3/2}$. \par
To assess whether this instability is accessible with realistic materials, we place three representative materials used in the experiments of ref.~\cite{kaneelil2026}: natural rubber, HDPE and a soft vinyl polysiloxane (VPS), with Young's moduli from $0.2$~MPa to $1.3$~GPa and thicknesses of $0.4$--$1.6$~mm (Table~S1). To eliminate the gravity effect, these sheets are assumed to be density-matched in water, and the calculated $K$ ranges from $0.05$ to $\sim40$. We numerically probe three different angular velocities $\Omega/2\pi=0.1$, $1$ and $10$~Hz. Each sheet sample occupies a vertical column with filled markers corresponding to unstable states and open markers corresponding to stable states. 

\section{Conclusions}
Rotation can both stabilize and destabilize a thin elastic disk immersed in a viscous fluid. By coupling the mechanics of a rotating elastic plate to the steady tractions generated by the surrounding swirling flow, we have shown how this competition gives rise to an elastohydrodynamic buckling instability. While centrifugal forces and radial viscous tractions generate stabilizing tensile stresses, azimuthal viscous tractions generate destabilizing shear stresses. The resulting stability problem is mathematically equivalent to that of a disk undergoing angular acceleration \cite{sader2019shear,coman2023shear}, with the important difference that the destabilizing shear is sustained even when the disk rotates at a constant angular velocity.  

The onset and morphology of buckling are governed, for a given Poisson ratio, by three dimensionless parameters: the radius ratio $\alpha$, and two dynamical dimensionless parameters, the solid number $So$, which measures the ratio of centrifugal tension to bending stiffness, and the fluid number $Fl$,  which measures viscous traction relative to centrifugal loading. 
We find that the flat disk buckles into an azimuthally periodic buckled shape when these parameters jointly exceed a critical threshold.  Two limiting regimes emerge: at small $So$, bending resists the fluid-induced stresses, whereas at large $So$, stability is governed primarily by the competition between shear and tension. The critical buckling modes exhibit a spiral morphology, with the azimuthal wavenumber increasing with both solid number $So$ and radius ratio $ \alpha$. As centrifugal tension increases, the deformation becomes increasingly confined to the vicinity of the clamped inner edge, where the shear stress is largest.

Perhaps the most striking consequence of this competition is the emergence of re-entrant stability. Since viscous stresses scale as $\Omega^{3/2}$ while centrifugal stresses scale as $\Omega^{2}$, the disk is stabilized by bending at low rotation rates and by centrifugal tension at high rotation rates, but may buckle at intermediate rotation rates. For a given disk–fluid system, the rotation-independent parameter $K=So^{1/2}Fl^2$  determines whether this instability can occur. Below a geometry-dependent critical value of $K$, the flat configuration remains stable against static buckling at all rotation rates; above it, buckling occurs over a finite interval of angular velocities. Estimates using representative elastic materials suggest that this phenomenon should be experimentally accessible.


Our analysis provides a minimal description of the instability, but also raises several questions. How does the finite geometry of the disk modify the swirling flow, particularly near the inner edge where the deformation tends to localize? How does the deformation of the disk, in turn, modify the flow and the resulting pressure and viscous tractions? Can this feedback alter the onset of static buckling or give rise to new dynamical instabilities? Beyond threshold, the nonlinear coupling between bending, stretching and fluid motion will determine the amplitude, morphology and dynamics of the buckled configurations. Incorporating gravity would further extend the theory to initially nonplanar rotating sheets and membranes   \citep{chen2011warping,delapierre2018wrinkling,guven2013whirling,kaneelil2026}.  

An immediate experimental test of our predictions would be to follow the onset, evolution and disappearance of buckling as the rotation rate of a density-matched elastic disk immersed in a viscous fluid is varied. More generally, the interplay between elastic resistance, inertial tension and fluid-induced shear provides a mechanism for controlling the shape of compliant rotating structures through rotation alone.

\section*{Acknowledgements}  
We would like to thank the anonymous reviewers for correcting a critical mistake in our earlier version of our manuscript. We acknowledge partial funding from the European Union's Horizon Europe research and innovation programme under the Marie Skłodowska-Curie grant agreement (No. 101203107) (S.Y.), as well as the Simons Foundation and the Henri Seydoux Fund (L.M.). 

\section*{Data availability statement} The linear stability analysis and numerical simulation codes are available on a GitHub repository https://github.com/YinSifan0204/Elastohydrodynamic-instability

\bibliography{Theory.bib}

\end{document}


\title{Supplemental Material: \\Buckling and restabilization of a spinning elastic disk in a viscous fluid}

\author{Sifan Yin}
\affiliation{ Max Planck Institute for the Physics of Complex Systems, 01187 Dresden, Germany.}
\author{Paul R. Kaneelil}
\affiliation{ John A. Paulson School of Engineering and Applied Sciences, Harvard University, Cambridge, MA 02138, USA.}

\author{L.\ Mahadevan}
 \email{lmahadev@g.harvard.edu}
\affiliation{John A. Paulson School of Engineering and Applied Sciences, Harvard University, Cambridge, MA 02138, USA}
\affiliation{Department of Organismic and Evolutionary Biology, Harvard University, Cambridge, MA 02138}
\affiliation{Department of Physics, Harvard University, Cambridge, MA 02138}

\maketitle

\section{Discussion of the Coriolis term}
Before buckling, the static displacement field of the disk spinning at a constant rate $\Omega$ is given by only in-plane deformation ${\bm u} = (u_r,u_\theta,0)^T$ which can be represented by the mid-plane displacement field ${\bm u}^{\rm mid}=(u,v,0)^T$, $u_r=u, u_\theta=v$. The body force comprises a centrifugal term ${\bm f}_{\rm centifugal} = \Omega^2 r{\bm e}_r$ and the shear ${\bm f}_{\rm fluid}$ induced by the swirling fluids. The mechanical equilibrium equation of the unbuckled disk reads
$\nabla\cdot{\bm \sigma}+{\bm f}_{\rm centifugal} + {\bm f}_{\rm fluid}=0$.

After buckling, the displacement field ${\bm u} = (u_r,u_\theta,u_z)^T$  of an arbitrary point at location $(r,\theta,z)$ and at time point $t$ can be expressed by the mid-plane displacement field ${\bm u}^{\rm mid}=(u,v,w)^T$
$u_r = u-z\frac{\partial w}{\partial r}$, $ u_\theta=v-\frac zr\frac{\partial w}{\partial \theta}$, $u_z=w$.
At the mid-plane, the radial and azimuthal displacements remain the same as they were before buckling and do not change with time. However, the normal displacement field $w$ of the rotating disk with a traveling mode undulation is a function of $(r,\theta,t)$ and would induce a Coriolis force. The relative velocity of any material point within the disk is
\begin{equation}
    {\bm v}_{\rm rel} = \frac{\partial {\bm u}}{\partial t}=-z\frac{\partial \dot w}{\partial r}{\bm e}_r-\frac zr\frac{\partial \dot w}{\partial \theta}{\bm e}_{\theta}+{\dot w}{\bm e}_{z}.
\end{equation}

The corresponding Coriolis force per unit volume is
\begin{equation}
\mathbf f_{\rm C}=-2\rho_s\Omega\,\mathbf e_z\times\mathbf v_{\rm rel}=2\rho_s\Omega\,z\,\mathbf e_z\times\nabla\dot w
=2\rho_s\Omega z\Big(-\frac1r\frac{\partial\dot w}{\partial\theta}\mathbf e_r+\frac{\partial\dot w}{\partial r}\mathbf e_\theta\Big).
\end{equation}
The transverse velocity $\dot w\,\mathbf e_z$ is parallel to the rotation axis and produces no Coriolis force; only the in-plane velocity associated with the rotation of the plate normals does. This force is in-plane and odd in $z$, so its resultant over the thickness vanishes, $\int_{-h/2}^{h/2}\mathbf f_{\rm C}\,dz=\mathbf 0$, and it does not alter the membrane equilibrium. Its first moment is a distributed couple per unit area,
\begin{equation*}
\mathbf m_{\rm C}=\int_{-h/2}^{h/2}z\,\mathbf f_{\rm C}\,dz=\frac{\rho_s\Omega h^3}{6}\,\mathbf e_z\times\nabla\dot w .
\end{equation*}
In Kirchhoff theory, such a couple modifies the transverse shear force, $\mathbf Q=\nabla\cdot\mathbf M+\mathbf m_{\rm C}$, and therefore enters the transverse equilibrium equation through $\nabla\cdot\mathbf m_{\rm C}$. However,
\begin{equation*}
\nabla\cdot\big(\mathbf e_z\times\nabla\dot w\big)=\frac1r\Big[\frac{\partial}{\partial r}\Big(-\frac{\partial\dot w}{\partial\theta}\Big)+\frac{\partial}{\partial\theta}\Big(\frac{\partial\dot w}{\partial r}\Big)\Big]=-\,\mathbf e_z\cdot\nabla\times\nabla\dot w\equiv0 .
\end{equation*}
The Coriolis force therefore makes no contribution to the linearized equation for $w$. Therefore the co-rotating frame, the transverse dynamics of a spinning Kirchhoff plate contains no gyroscopic term, in agreement with the classical theory of spinning disks~\cite{lamb1921vibrations}. Gyroscopic coupling appears only when the motion is described in a space-fixed frame, through the convective derivative $\partial_t+\Omega\,\partial_\theta$. Similarly, the centrifugal force acting on the in-plane displacement $-z\nabla w$ of the off-midplane material produces a couple of order $\rho_s\Omega^2h^3\nabla w$, which is smaller than the retained membrane term $h\nabla\cdot(\boldsymbol\sigma\cdot\nabla w)$, with $\sigma\sim\rho_s\Omega^2R^2$, by a factor $(h/R)^2$, and is neglected together with rotary inertia. The governing equation for the normal deflection is thus Eq.~(1) of the main text, without any Coriolis contribution.


\section{Complete linear eigenvalue problem}
\subsection{Governing equations}
The in-plane force balance equations in the main text Eq.~(3) are expressed as
\begin{subequations}\label{Eq_in_plane_force_balance}
    \begin{align}
        h\frac{{\rm d}\sigma_{rr}}{{\rm d}r}+h\frac{\sigma_{rr}-\sigma_{\theta\theta}}{r}+2T_{r}+h\rho_s\Omega^2 r&=0,\\
        h\frac{{\rm d}\sigma_{r\theta}}{ {\rm  d} r}+h\frac{2\sigma_{r\theta}}{r}+2T_\theta&=0,
    \end{align}
\end{subequations}
where 
\begin{equation}
     T_r = F'\left(0\right)\rho_f(\Omega^3\nu)^{1/2}r,
     ~{\rm and}~
     T_{\theta} = -G'\left(0\right) \rho_f(\Omega^3\nu)^{1/2}r,
\end{equation}
with $G'\left(0\right) \approx 0.616$ and $F'\left(0\right)\approx 0.510$. Non-dimensionalizing the governing equation by defining the dimensionless variables $\tilde{r} = r/R$, $\varepsilon=h/R$, and $\pmb{\tilde {\sigma}}=\pmb{\sigma}/[\rho_s\Omega^2 R^2]$, we obtain
\begin{subequations}
    \begin{align}
        \frac{{\rm d}\tilde\sigma_{rr}}{{\rm d}\tilde r}+\frac{\tilde\sigma_{rr}-\tilde\sigma_{\theta\theta}}{\tilde r}+\left[1+2F'(0)Fl\right]\tilde r&=0,\\
        \frac{{\rm d}\tilde \sigma_{r\theta}}{ {\rm  d}\tilde r}+\frac{2\tilde \sigma_{r\theta}}{\tilde r}-2G'(0)Fl{\tilde r}&=0,
    \end{align}
\end{subequations}
The nondimensionalized in-plane stresses are solved as
\begin{equation}\label{Eq_pre_stress}
    \begin{aligned}
        \tilde \sigma_{rr}&=\left(1+2F'\left(0\right) Fl\right)\sigma_{rr}^{\rm c},~~\sigma_{rr}^{\rm c}=\left(-A_1{\tilde r}^2+\frac{B_1}{\tilde r^2}+C_1\right),\\
        {\tilde \sigma}_{\theta\theta}&=\left(1+2F'\left(0\right) Fl\right)\sigma_{\theta\theta}^{\rm c},~~\sigma_{\theta\theta}^{\rm c}=\left(-A_2{\tilde r}^2-\frac{B_1}{\tilde r^2}+C_1\right),\\
        {\tilde \sigma}_{r\theta}&=-\frac{G'\left(0\right)}{2} Fl\frac{1-\tilde r^4}{\tilde r^2}.
    \end{aligned}
\end{equation}

The coefficients are given as 
\begin{equation}
    \begin{aligned}
        A_1& = \frac{3+\nu_{\rm p}}{8},\\
        B_1& =\frac{\alpha^2\left(1-\nu_{\rm p}\right)\left[\left(3+\nu_p\right)-\alpha^2\left(1+\nu_{\rm p}\right)\right]}{8\left[1+\nu_{\rm p}+\alpha^2\left(1-\nu_{\rm p}\right)\right]},\\
         C_1& =\frac{\left(1+\nu_{\rm p}\right)\left[\left(3+\nu_p\right)+\alpha^4\left(1-\nu_{\rm p}\right)\right]}{8\left[1+\nu_{\rm p}+\alpha^2\left(1-\nu_{\rm p}\right)\right]},\\
         A_2& = \frac{1+3\nu_{\rm p}}{8},
    \end{aligned}
\end{equation}\label{eq:ABC_coeffients}
where $A_1=B_1+C_1$ and $3A_1-A_2=1$. Here, we find that the rotating disk at a constant angular velocity without any external loads would only have tensile stresses, which stabilize the disk. For axisymmetric stress distribution\par
\begin{equation}
    \nabla\cdot(\boldsymbol\sigma\cdot\nabla w)=\sigma_{rr}w_{rr}+\Big(\sigma_{rr}'+\frac{\sigma_{rr}}{r}\Big)w_r+\frac{\sigma_{\theta\theta}}{r^2}w_{\theta\theta}+\frac{2\sigma_{r\theta}}{r}w_{r\theta}+\frac{\sigma_{r\theta}'}{r}w_\theta.
\end{equation}

Setting the inertia term to be zero in main text Eq. (8), we obtain the equation for the onset of instability,
\begin{equation}\label{Eq_w_nondimen_equation}
\begin{aligned}  
   &
    \frac{\partial^4 w}{\partial r^4}+ \frac{2}{r}\frac{\partial^3 w}{\partial r^3} +\frac{1}{r^2}\bigg(2\frac{\partial^4 w}{\partial r^2 \partial \theta^2} -\frac{\partial^2w}{\partial r^2}\bigg) +
    \frac{1}{r^3}\bigg( \frac{\partial w}{\partial r} -2\frac{\partial^3 w}{\partial r \partial \theta^2}  \bigg)  +\frac{1}{r^4}\bigg(4\frac{\partial^2w}{\partial \theta^2}+\frac{\partial^4 w}{\partial \theta^4}\bigg) \\ 
    &=So\left(1+2F'(0)Fl\right) \left[ \left(-A_1r^2+\frac{B_1}{r^{2}}+C_1\right)\frac{\partial^2w}{\partial r^2} + \left(-3A_1r-\frac{B_1}{r^{3}}+\frac{C_1}{r}\right) \frac{\partial w}{\partial r} + \left(-A_2-\frac{B_1}{r^{4}}+\frac{C_1}{r^{2}}\right)\frac{\partial^2w}{\partial\theta^2}\right] \\ 
    &+SoG'(0)Fl\left(\frac{r^4-1}{r^3}\frac{\partial^2 w}{\partial r\partial \theta}+\frac{r^4+1}{r^4}\frac{\partial w}{\partial \theta}\right),
\end{aligned}
\end{equation}

Compared to Eq.(11) in \cite{sader2019shear}, we have the effective angular velocity and angular acceleration
\begin{subequations}\label{Eq_Comp_Sader}
    \begin{align}
        \bar\omega^2 &= So\left[1+2F'(0)Fl\right],\\
        \bar\Omega&=2SoG'(0)Fl.
    \end{align}
\end{subequations}
When there is no fluid environment, i.e., $Fl=0$, the soft disk rotates at a constant angular velocity, $\bar\omega^2=So$, and $\bar\Omega=0$.

The governing ordinary differential equation for $\phi(r)$ is
 \begin{equation}\label{Eq_phi_ODE_SI}
   0= -\phi^{(4)}(r) + c_3(r) \phi^{(3)}(r) + c_2(r) \phi^{(2)}(r) + c_1(r) \phi^{(1)}(r) + c_0(r) \phi(r),
\end{equation}
with coefficients
\begin{equation}\label{Eq_coeffs03}
\begin{aligned}
    c_3 &= -\frac{2}{r}, \\
    ~c_2 &= \frac{2m^2+1}{r^2}+So \left[1+2F'(0)Fl\right] \left(-A_1r^2+\frac{B_1}{r^2}+C_1\right) ,\\
    c_1 &= -\frac{2m^2+1}{r^3}+So\left[1+2F'(0)Fl\right] \left(-3A_1r-\frac{B_1}{r^3}+\frac{C_1}{r}\right) + {i}mSoG'(0)Fl \left(\frac{r^4-1}{r^3} \right), \\
    ~c_0 &= \frac{4m^2-m^4}{r^4}- So\left[1+2F'(0)Fl\right]\left(-A_2-\frac{B_1}{r^4}+\frac{C_1}{r^2}\right)m^2 + {i}mSoG'(0)Fl \left(\frac{r^4+1}{r^4} \right).   
    \end{aligned}
\end{equation}

The coefficients in the boundary condition in main text Eq.(15) are defined as
\begin{equation}
\begin{aligned}
    b_1 &= \nu_p,~b_0 = -\nu_p m^2 \\
    a_2 &= 1,~a_1 = m^2(\nu_p-2)-1,~a_0=m^2(3-\nu_p).
    \end{aligned}
\end{equation}
\subsection{Linear stability analysis and simulations of steady buckling states}
We rewrite Eq.~(\ref{Eq_phi_ODE_SI}) as a first-order system 
\begin{equation}
    \frac{{\rm d}{\bm y}}{{\rm d} r}={\bm A}(r)\cdot{\bm y}(r),
\end{equation}
where ${\bm y} = (y_1,y_2,y_3,y_4)^{\rm T}$,
\begin{equation}
    {\bm A}(r) = \begin{pmatrix}
        0 & 1 & 0 & 0\\
        0 & 0 & 1 & 0\\
        0 & 0 & 0 & 1\\
        c_0 & c_1 & c_2 & c_3
    \end{pmatrix}
\end{equation}
We use a determinant method \citep{haughton1979bifurcation,zhu2008asymmetric,vandiver2009differential} and the compound matrix method to solve this BVP eigenvalue problem. 
For different $m$, we find the smallest $Fl$ as the bifurcation threshold. The critical $Fl$ as a function of $So$ is presented in Fig. 2(a) in the main text.  

\section{Comparison to former work \cite{sader2019shear,coman2023shear}}
\subsection{Comparison to spinning-up disk in a vacuum}
\begin{figure}
\centering
\includegraphics[width=\textwidth]{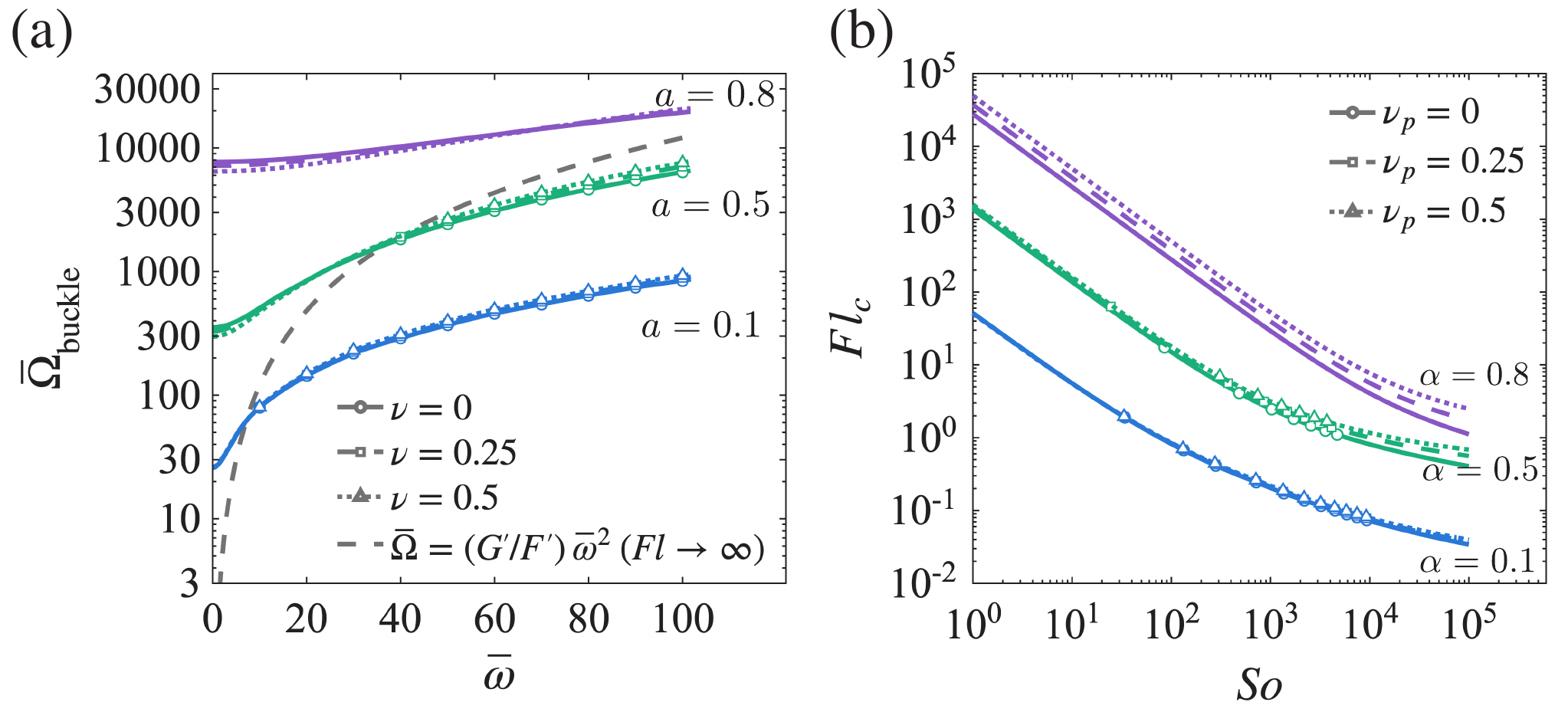}
\caption{Comparison with \citet{sader2019shear}. (a) Critical angular acceleration $\bar\Omega_{\rm buckle}$ versus angular velocity $\bar\omega$. Lines are data of Fig.~2(C) of \citet{sader2019shear} and  markers are effective $\bar \Omega$ and $\bar \omega$ computed from our model Eq.~\ref{Eq_Comp_Sader}. Dashed curve: $\bar\Omega=(G'(0)/F'(0))\,\bar\omega^2$, corresponding to $Fl\to\infty$, above which points are not reachable for $Fl>0$. (b) Critical fluid number $Fl_{\rm c}$ as a function of $So$. Lines, our model; markers, data of \citet{sader2019shear} mapped through Eq.~\ref{Eq_Comp_Sader}.}
\label{figS_compare_Sader2019}
\end{figure}

The eigenvalue problem Eq.~（\ref{Eq_w_nondimen_equation}） has the same form as Eq.~(11) of \citet{sader2019shear} for a clamped–free annular disk undergoing spin-up, with the angular velocity $\bar\omega$ and angular acceleration $\bar\Omega$ replaced by our Eq.~\ref{Eq_Comp_Sader}. The radial drag thus increases the effective centrifugal tension, while the azimuthal drag plays the role of an effective angular acceleration. Fig.~\ref{figS_compare_Sader2019}(a) compares the effective critical acceleration $\bar\Omega_{\rm buckle}(\bar\omega)$ (markers), computed from Eq. ~\ref{Eq_Comp_Sader}(b), with the data of Fig.~2(C) of \citet{sader2019shear} (lines), for $a=0.1$, $0.5$, $0.8$ and $\nu=0$, $1/4$, $1/2$. The maximum relative difference is below $0.4\%$, at the level of the accuracy with which the published curves can be read. Conversely, in Fig.~S\ref{figS_compare_Sader2019}(b) the data of Fig.~2(C) of \citet{sader2019shear} (markers) are mapped to $(So,Fl)$ and lie on our critical curves $Fl_{\rm c}(So)$. In our model, there is a limit 
\begin{equation}
\frac{\bar\Omega}{\bar\omega^2}=\frac{2G'(0)Fl}{1+2F'(0)Fl}<\frac{G'(0)}{F'(0)}\approx1.21.
\end{equation}
Therefore, only part of Sader's parameter space corresponds to out model. Points above the dashed curve $\bar\Omega=(G'/F')\,\bar\omega^2$ in panel (a) cannot be reached by a disk spinning in a viscous fluid. For $a=0.8$ this excludes the whole range $\bar\omega\le100$ shown by \citet{sader2019shear}, so that no mapped points appear in panel (b) for $\alpha=0.8$.   

\subsection{Membrane limit $So\to\infty$}\label{SM_membrane}
For $So\to\infty$， the bending term in Eq.~\ref{Eq_w_nondimen_equation} drops out and the stability problem reduces to $\nabla\cdot(\boldsymbol\sigma\cdot\nabla w)=0$. Stability therefore depends only on the prestress Eq.~\ref{Eq_pre_stress}. We define the ratio of shear to tension as 
\begin{equation}
\lambda=\frac{G'(0)Fl}{1+2F'(0)Fl}<\frac{G'(0)}{2F'(0)}\approx0.60.
\end{equation}
A membrane under tensile principal stresses cannot buckle. Since $\sigma_{rr},\sigma_{\theta\theta}\ge0$ in the annulus, a compressive principal stress appears when $\sigma_{r\theta}^2>\sigma_{rr}\sigma_{\theta\theta}$, which requires
\begin{equation}
\lambda>\lambda_{\rm b}(r)\equiv\frac{2\sqrt{\tilde\sigma_{rr}^{\rm c}(r) \tilde\sigma_{\theta\theta}^{\rm c}(r)}}{r^{-2}-r^2}.
\end{equation}
Because the shear stress is largest at the clamped edge, $\lambda_{\rm b}(r)$ is minimal at $r=\alpha$, and compression first appears there when $\lambda$ exceeds
\begin{equation}
\lambda_{\rm b}(\alpha)=\frac{2\sqrt{\tilde\sigma_{rr}^{\rm c}(\alpha)\tilde\sigma_{\theta\theta}^{\rm c}(\alpha)}}{\alpha^{-2}-\alpha^2}.
\end{equation}
Without bending stiffness, wrinkles of arbitrarily short wavelength can form in an arbitrarily thin compressive region, so that the critical ratio $\lambda_{\rm c}\to\lambda_{\rm b}(\alpha)$ as $So\to\infty$; for finite $So$, $\lambda_{\rm c}>\lambda_{\rm b}(\alpha)$, in agreement with the lower bound of \citet{coman2023shear}, whose shear-to-tension ratio equals $\lambda/2$. Setting $\lambda(Fl_{\rm c}^{\rm lim})=\lambda_{\rm b}(\alpha)$ gives
\begin{equation}
Fl_{\rm c}^{\rm lim}=\frac{\lambda_{\rm b}(\alpha)}{G'(0)-2F'(0)\lambda_{\rm b}(\alpha)},
\end{equation}
which exists provided $\lambda_{\rm b}(\alpha)<G'(0)/[2F'(0)]$. For $\nu_p=1/4$, $\lambda_{\rm b}(\alpha)=0.0064$, $0.052$, $0.120$ and $Fl_{\rm c}^{\rm lim}=0.011$, $0.092$, $0.24$ for $\alpha=0.1$, $0.3$, $0.5$. 

\section{Experimental feasibility: geometry and material parameters}
Here we choose three representative materials for the taken from the experiments of ref.~\cite{Kaneelil2026}: a soft vinyl polysiloxane (VPS), natural rubber and high-density polyethylene (HDPE). Their Young's moduli range from $0.2$~MPa to $1.3$~GPa, and we use the thicknesses of those experiments (Table~S1). The sheets are clamped at $R_i=1.6$~cm and we consider $\alpha=0.1$, $0.2$ and $0.3$, i.e. $R=R_i/\alpha=16$, $8$ and $5.3$~cm. These sheets are assumed to be density-matched with water ($\rho_f=\rho_s$). For each sheet, the rotation-independent parameter $K=So^{1/2}Fl^2=$ is calculated directly from the material and geometric parameters. In Fig.~3 in the main text, each sheet rotates at $\Omega/2\pi=0.1$, $1$ and $10$~Hz, which covers $10^{-4}\lesssim So\lesssim5\times10^{5}$ and $0.08\lesssim Fl\lesssim3$. The thin-plate and thin-boundary-layer assumptions hold throughout, with $h/R\leq0.03$ and $\delta/R\leq0.08$. The critical curves are computed as $\nu_p=0.5$ for VPS and rubber, and $\nu_p=0.46$ for HDPE.

\begin{table}[t]
\caption{Parameters of the sheets of Ref.~\cite{Kaneelil2026} used in main text Fig.~3, and the calculated parameter $K$ (Eq.~(16) in the main text). The disks are density-matched with water ($\rho_f=\rho_s$,
$\nu=1.0\times10^{-6}$~m$^2$\,s$^{-1}$) and clamped at $R_i=1.6$~cm. Each sheet rotates at $\Omega/2\pi=0.1$, $1$, $10$~Hz.}
\label{Table_materials}
\begin{ruledtabular}
\begin{tabular}{lcccccc}
 & & & & \multicolumn{3}{c}{$K$}\\
\cline{5-7}
Material & $E$ & $\nu_p$ & $h$ (mm) & $\alpha=0.1$ & $\alpha=0.2$ & $\alpha=0.3$\\
\hline
VPS    & $0.2$~MPa & $0.50$ & $0.50$ & $43.40$   & $10.85$    & $4.82$\\
VPS    & $0.2$~MPa & $0.50$ & $0.65$ & $19.75$   & $4.94$   & $2.20$\\
VPS    & $0.2$~MPa & $0.50$ & $1.30$ & $2.47$  & $0.62$  & $0.27$\\
Rubber & $1.9$~MPa & $0.50$ & $1.57$ & $0.45$ & $0.11$  & $0.05$\\
HDPE   & $1.3$~GPa & $0.46$ & $0.42$ & $0.93$ & $0.23$  & $0.10$\\
\end{tabular}
\end{ruledtabular}
\end{table}

\bibliographystyle{jfm}
\bibliography{Theory.bib}